\documentclass[numberedappendix]{emulateapj}
\usepackage{amsbsy,wasysym, color}
\usepackage{times}

\renewcommand{\b}{\hat{\bb{b}}}
\newcommand{\na}{\bb{\nabla}}
\newcommand{\be}{\begin{eqnarray}}
\newcommand{\en}{\end{eqnarray}}
\newcommand{\pa}{\partial}
\newcommand{\f}{\frac}
\newcommand{\tci}{\tau^{-1}_{\rm c}}
\newcommand{\tdi}{\tau^{-1}_{\rm d}}
\newcommand{\tvi}{\tau^{-1}_{\rm v}}

\newcommand\bb[1]{\mbox{\boldmath{$#1$}}}
 
\newcommand\bcdot{\bb{\cdot}}
\newcommand\btimes{\bb{\times}}
\begin{document}

\title{The Stability of Dilute Plasmas with Thermal and Composition Gradients. I. \\
The Slow Conduction Limit: Overstable Gravity Modes}

\author{Martin~ E. Pessah and Sagar Chakraborty\footnote{mpessah@nbi.dk; sagar@nbi.dk}}
\affil{
Niels Bohr International Academy, Niels Bohr Institute, Blegdamsvej 17, 2100 Copenhagen \O, Denmark}

\shorttitle{Dilute Plasmas with Composition Gradients. I.}
\shortauthors{Pessah \& Chakraborty}

\begin{abstract}
We analyze the stability of a dilute plasma with thermal and composition gradients
in the limit where conduction is slow compared to the dynamical timescale. 
We find necessary and sufficient conditions for stability 
when the background magnetic field is either parallel or perpendicular to the 
thermal and composition gradients that are parallel to the gravitational field.
We provide approximate solutions for all the relevant modes involved, which are driven by
gravity, conduction, and diffusion. We discuss the astrophysical implications of our findings
for a representative  galaxy cluster where helium has sedimented.
\end{abstract}


\section{Introduction}
\label{sec:introduction}

Despite the fact that magnetic fields in galaxy clusters are too weak
to be mechanically important, they can play a fundamental role in the
dynamical stability of the dilute gas by channelling the transport of
heat. The collision-less character of the hot intra-cluster medium
(ICM), which is generically characterized by stable entropy gradients
according to Schwarzschild's criterion \citep{2005A&A...433..101P,
2009ApJS..182...12C}, enables the action of magnetic instabilities
that are sensitive to temperature gradients
\citep{2000ApJ...534..420B, 2004ApJ...616..857B}.  In particular, the
magneto-thermal instability (MTI) operates when magnetic field lines
are orthogonal to a temperature gradient parallel to the
gravitational field \citep{2001ApJ...562..909B}, whereas the
heat-flux-driven buoyancy instability (HBI) acts when magnetic field
lines are parallel to a temperature gradient anti-parallel to the
gravitational field \citep{2008ApJ...673..758Q}. Numerical simulations
suggest that the non-linear evolution of these instabilities leads to
magnetic field configurations that tend to suppress the ability of the
plasma to tap into the free energy supplied by the background
temperature gradient \citep{2009ApJ...703...96P, 2009ApJ...704..211B,
2010ApJ...713.1332R}.  These magnetic field configurations can
nevertheless support overstable gravity modes
\citep{2010ApJ...720L..97B}.

While the landscape of thermal instabilities that render homogeneous,
dilute plasmas unstable has been well explored
\citep{2011MNRAS.417..602K}, and even extended to account for the effects of cosmic rays
\citep{2006ApJ...642..140C}, very little is known about the effects
that composition gradients can have on the stability of the dilute
ICM.  If magnetic fields do not prevent the efficient diffusion of
ions \citep{2001ApJ...562L.129N, 2003MNRAS.342L...5C,
2004MNRAS.349L..13C} then the gradients in mean molecular weight can be as
important as the gradients in temperature (see Section~\ref{sec:discussion} and
\citealt{2000ApJ...529L...1Q,
2009ApJ...693..839P, 2010MNRAS.401.1360S, 2011A&A...533A...6B}) and
provide another source of free energy to feed instabilities.
In order to obtain a more complete picture of the stability
properties of the ICM, it is thus important to relax the
assumption of a homogeneous plasma.

As a first step towards understanding the role of composition
gradients in dilute plasmas, where magnetic fields play a key role by
channelling the conduction of heat and the diffusion of ions, we
investigate the stability of modes for which heat conduction is slow
compared to the dynamical timescale involved. These modes contain,
among others, the overstable $g$-modes studied by
\citet{2010ApJ...720L..97B}.

\section{Model for the Multicomponent, Dilute Atmosphere}
\label{sec:model}

\subsection{General Considerations for the Plasma Model}

In order to highlight the physical phenomena that emerge when
composition gradients are accounted for, we focus our attention on a
dilute binary mixture (e.g., hydrogen and helium)\footnote{It is straightforward
  to generalize the present analysis to $N$ species.} in a fixed
gravitational field described by\footnote{For the sake of simplicity, we do not consider relatively weaker
effects, such as  thermo-difussion, baro-difussion, etc. \citep{1959flme.book.....L}.} 
\be &&\f{\pa \rho}{\pa
  t}+\bb{\nabla}\bcdot(\rho \bb{v})=0 \,,
\label{rho}\\
&&\f{d \bb{v}}{dt} =-\f{1}{\rho}\bb{\na} \bcdot \left(\mathsf{P} + 
\f{\bb{B}^2}{8\pi}\mathsf{I} - \f{{B^2}}{4\pi}\hat{\bb{b}}\hat{\bb{b}}\right) + \bb{g} \,, \,\,\,\,
\label{v}\\
&&\f{\pa \bb{B}}{\pa t}=\bb{\na}\btimes(\bb{v}\btimes\bb{B}) \,,
\label{b}\\
&& \f{P}{\gamma-1} \f{d}{dt}(\ln{P\rho^{-\gamma}})=
(p_\bot-p_\parallel)\f{d}{dt}\ln\f{B}{\rho^{\gamma-1}} 
- \na.\bb{Q}_{\rm s} 
 \,, \,\,\,\,\,\,\,
\label{S} \\
&&\f{dc}{dt} =-\na\bcdot\bb{Q}_{\rm c} \,.
\label{c}
\en 
Here $\rho$ is the mass density, $\bb{v}$ is the fluid velocity,
$\bb{B}$ is the magnetic field, $\bb{g}$ is the gravitational
acceleration, and $\gamma$ is the adiabatic index. The Lagrangian and
Eulerian derivatives are related via $d/dt \equiv \pa/\pa t + \bb{v}
\bcdot \nabla$.

Equations~(\ref{rho})--(\ref{c}) describe the dynamics of a
binary mixture in the low-collisionality regime and they differ from
standard MHD in three important respects:

(\emph{i}) The pressure tensor $\mathsf{P} \equiv p_\bot \mathsf{I} +
(p_\parallel - p_\bot) \hat{\bb{b}} \hat{\bb{b}}$, is anisotropic;
where the symbols $\bot$ and $\parallel$ refer to the directions
perpendicular and parallel to the magnetic field, whose direction is
given by the versor $\hat{\bb{b}}\equiv \bb{B}/B$. 

(\emph{ii}) Heat flows mainly along magnetic field lines,
because the electron mean free path is large compared to its
Larmor radius. This process is modeled by the second term on the right
hand side of Equation~(\ref{S}) via $\bb{Q}_{\rm s}
\equiv-\chi\b(\b\bcdot\na)T$, where $T$ is the plasma temperature,
assumed to be the same for ions and electrons, and $\chi \approx 6
\times 10^{-7} T^{5/2}$ ergs cm$^{-1}$ s$^{-1}$ K$^{-1}$ is the
thermal conductivity \citep{1962pfig.book.....S, 1965RvPP....1..205B}.

(\emph{iii}) The composition of fluid elements can change due to
particle fluxes.  Considering $\bb{Q}_{\rm c}\equiv-D\b(\b\bcdot\na)c$
on the right hand side of Equation~(\ref{c}) ensures that the diffusion
of ions is mainly along magnetic field lines.  This is a good
approximation when the plasma is dilute enough for the ion mean free
path to be large compared to the ion Larmor radius. Note that the
concentration $c$ is related to the mean molecular weight via 
$1/\mu \equiv (1-c)(1+Z_1)/\mu_1 + c(1+Z_2)/\mu_2$, 
where $\mu_i$ and $Z_i$, with $i=1,2$, are the molecular weights 
and the atomic numbers for the two ion species. The isotropic
part of the pressure tensor is thus $P\equiv 2p_\bot/3 + p_\parallel/3 = \rho k_{\rm B}
T/\mu m_{\rm H}$, where $k_{\rm B}$ is the Boltzmann constant and 
$m_{\rm H}$ is the atomic mass unit.

\begin{figure*}[t]
\begin{center}
  \includegraphics[width=\textwidth,trim=0 0 0 0]{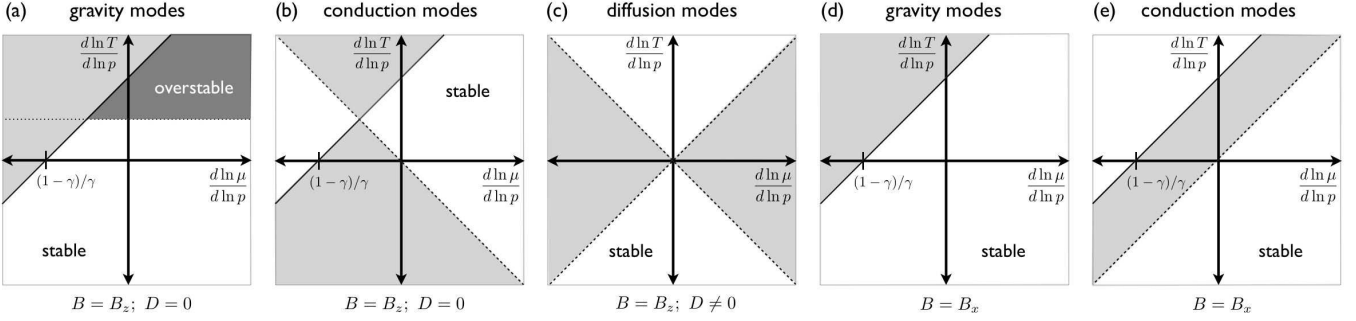}
  \caption{Graphic representation of the stability of modes for which 
  conduction is slow compared to the dynamical timescale, i.e., $\omega_{\rm dyn} \gg \tci$.
The various panels show the unstable regions (gray) for each of the modes that can be excited
when the background magnetic field is parallel ($a$, $b$, and $c$)
or perpendicular ($d$ and $e$) to the background thermal and composition gradients.
The solid line corresponds to $N^2=0$; 
the horizontal dotted line represents $d\ln T/d\ln P = (\gamma-1)/2\gamma$; and
the dashed lines correspond to $d\ln T/d\ln P = \pm  d\ln \mu/d\ln P$.
For $B= B_z$, gravity modes ($a$) can become either stable or overstable, 
while modes driven by conduction can become unstable ($b$). If ions can 
diffuse efficiently along magnetic field lines, a new type of mode can become unstable ($c$).
For $B= B_x$, both gravity modes ($d$) and conduction modes
can become unstable ($e$); while ion diffusion only leads to decaying modes.
  }
  \label{fig:f1}
\end{center}
\end{figure*}

\subsection{Background State and Equations for the Perturbations}

We assume a plane-parallel atmosphere in a 
constant gravitational field $\bb{g}\equiv-g\hat{\bb{z}}$ which is
stratified in both temperature and composition along the vertical
direction. We consider a background magnetic field which is weak
enough that the mechanical equilibrium of the atmosphere, with scaleheight $H$,
is maintained via $dP/dz=-g\rho$.  In general, the background heat and particle
fluxes do not vanish, i.e., $\hat{\bb{b}}\bcdot \na T \ne 0$ and
$\hat{\bb{b}}\bcdot \na c \ne 0$, unless the magnetic field and the
background gradients are orthogonal. The existence of a well defined steady state, 
i.e., $\na\bcdot\bb{Q}_{\rm s}=\na\bcdot\bb{Q}_{\rm c}=0$,
demands that the background gradients should be linear functions 
of the distance along the direction of the magnetic field. However, even if this condition 
is not strictly satisfied, the dynamics of the modes that we consider is unlikely 
to be significantly affected if the local dynamical timescale is short compared 
to the timescale in which the entire system evolves (see also \citealt{2008ApJ...673..758Q}).

The modes of interest have associated timescales that are long
compared to the sound crossing time and it thus suffices to work in
the Boussinesq approximation. In this limit, the equations for the
linear  perturbations $\delta \sim e^{\sigma t + i \bb{k}\bcdot\bb{x}}$ become 
\be
\sigma\delta \bb{v}&=&-g\f{\delta\rho}{\rho}\hat{\bb{z}}-i\bb{k}v_{\rm th}^2\left(\f{\delta p_\bot}{P}+
\f{1}{\beta}\f{\delta B_\parallel}{B}\right) +ik_\parallel v_{\rm A}^2\f{\delta\bb{B}}{B}\nonumber\\&&-\b\f{3k_\parallel^2v^2_{\rm th}}{2\nu}\delta v_\parallel \,, \\
\label{deltav}
\sigma\delta \bb{B}&=&ik_\parallel B\delta \bb{v} \,, \\
\sigma\f{\delta\rho}{\rho}&=&\f{N^2}{g}\delta v_z +\frac{\gamma-1}{\gamma} \kappa k_\parallel^2\f{\delta T}{T}\nonumber\\
&&-i\frac{\gamma-1}{\gamma}\kappa\bb{k}\bcdot\left(\f{d\ln T}{dz}\delta b_z\b+ b_z\f{d\ln T}{dz}\f{\delta\bb{B}_\bot}{B}\right) \,,\\
\label{deltaT}
\sigma\f{\delta\mu}{\mu}&=&-\f{d\ln\mu}{dz}\delta v_z-Dk_\parallel^2\f{\delta \mu}{\mu}\nonumber\\
&&+iD\bb{k}\bcdot\left(\f{d\ln\mu}{dz}\delta b_z\b+b_z\f{d\ln\mu}{dz}\f{\delta\bb{B}_\bot}{B}\right)\,.
\label{deltamu}
\en 
In agreement with the Boussinesq approximation, the velocity
perturbations satisfy $\bb{k}\bcdot \delta \bb{v} = 0$ and the
fluctuations in density, temperature, and mean molecular weight are
related via 
\be
\f{\delta \rho}{\rho}+\f{\delta T}{T}-\f{\delta\mu}{\mu}=0 \,.
\label{d_rho_T_mu_eq0}
\en
Here, we have introduced the Alfv{\'e}n speed, $\bb{v}_{\rm A}\equiv \bb{B}/\sqrt{4\pi\rho}$, the
thermal speed, $v_{\rm th}\equiv \sqrt{2P/\rho}$, 
the plasma $\beta\equiv v_{\rm th}^2/v_{\rm A}^2$, 
the viscosity $\nu$ of the binary mixture, the 
thermal diffusion coefficient
$\kappa\equiv \chi T/P$,  and the Brunt$-$V{\"a}is{\"a}l{\"a} frequency
\be
N^2 \equiv \frac{g}{\gamma} \f{d}{dz}\ln P\rho^{-\gamma} = 
g \f{d}{dz}\ln \left(\f{P^{\f{1-\gamma}{\gamma}}T}{\mu} \right)\,.
\en

For completeness, we define here several quantities that play an
important role in the stability analysis. We denote by
$\omega_{\rm dyn} \equiv (g/H)^{1/2}$ and $\omega_{\rm A} \equiv \bb{k} \bcdot \bb{v}_{\rm A}$
the dynamical and Alfv{\'e}n frequencies, respectively.
The inverse time-scales
$\tci\equiv \kappa (\bb{k} \bcdot \hat{\bb{b}})^2(\gamma-1)/\gamma$, $\tdi\equiv D
(\bb{k} \bcdot \hat{\bb{b}})^2$, and $\tvi\equiv  3 k_\parallel^2 v_{\rm th}^2/2\nu$ 
characterize, respectively, the diffusion of heat, particles, 
and momentum along magnetic field lines. We also define two quantities 
which appear naturally when thermal and composition gradients are considered
\be N_{T\mu}^2 \equiv g\f{d}{dz}\ln
(T\mu)\,, \quad N_{T/\mu}^2 \equiv g\f{d}{dz}\ln
\left(\f{T}{\mu}\right) \,. \en

\section{Linear Mode Analysis}
\label{sec:modes}

We are concerned with  modes for which heat conduction is slow compared to the dynamical 
timescale and thus $\omega_{\rm dyn} \gg \tci$. 
Because the timescales associated with viscous processes, which are comparable 
to those involved in diffusion, are longer than the conduction timescales
by an order of magnitude \citep{2011MNRAS.417..602K}, we 
also have $\tci > \tvi \sim \tdi$.
We further focus our attention on modes for which magnetic tension is unimportant, i.e., 
$k_\parallel H\ll \beta^{1/2}$, and thus the Alfv{\'e}n frequency is small compared to
other inverse timescales. Since the plasma $\beta$ ranges from $10^2$ from 
the centers of cool core clusters to $10^4$ in the outskirts of the ICM
\citep{2002ARA&A..40..319C}, there is 
a reasonable range of wavenumbers for which a local analysis is sensible.
We can thus summarize the regime in which we are interested according to 
\be
\omega_{\rm dyn} \gg \tci > \tvi \sim \tdi \gg \omega_{\rm A} \,.
\en
For a homogeneous plasma, the modes satisfying these conditions encompass the 
overstable $g$-modes studied in \citet{2010ApJ...720L..97B}.
The set of Equations~(\ref{deltav})--(\ref{d_rho_T_mu_eq0})
allows us to address the behavior of these, as well as other new modes, 
in the presence of a non-vanishing gradient in the mean molecular weight
and account self-consistently for the diffusion of ions along magnetic field lines.

\subsection{Heat Diffusion Along Vertical Magnetic Field Lines with No-Ion Diffusion}
\label{sec:Bz_Deq0}

Let us first consider the case in which there is no ion diffusion.  
Setting $D=\nu=0$ the dispersion relation corresponding to
Equations~(\ref{deltav}) to (\ref{deltamu}) factorizes and possesses
as non-trivial solutions two Alfv{\'e}n waves, with $\sigma = \pm
i\omega_{\rm A}$, and the roots of the polynomial 
\be
\sigma^3+a_1\sigma^2+a_2\sigma+a_3=0 \,,
\label{pol_a}
\en 
with coefficients
\be
&&a_1 = \tci \,, \\
\label{a1}
&&a_2 =  \f{k_\bot^2}{k^2} N^2+\omega_{\rm A}^2 \approx  \f{k_\bot^2}{k^2} N^2 \,, \\
&&a_3 = -\tci \left(\f{k_\bot^2}{k^2} N_{T\mu}^2 - \omega_{\rm
    A}^2\right) \approx -\tci \f{k_\bot^2}{k^2} N_{T\mu}^2 \,,
\label{a3}
\en
where the approximate expressions for the coefficients follow from the
general considerations outlined above.

The Routh--Hurwitz stability criteria that predict exclusively
negative real parts for the roots of the cubic Equation~(\ref{pol_a}) are: $a_1>0,
a_3>0,$ and $a_1a_2-a_3>0$. The first condition is trivially satisfied
while the other two imply
\be
\label{RH-3.2}
N_{T\mu}^2 < 0 \,,\\
\label{RH-3.3}
N^2 +N_{T\mu}^2 > 0 \,,
\en
respectively. To leading order, two of the solutions of Equation~(\ref{pol_a})
are given by
$\sigma\approx\pm i a_2^{1/2}+(a_3-a_1a_2)/2a_2$, i.e.,
\be
\sigma&\approx&\pm i\f{k_\bot}{k}\sqrt{N^2} - \f{1}{2\tau_{\rm c}} \left(1 + \f{N_{T\mu}^2}{N^2}\right) \,,
\label{fast_overstable_Bz}
\en
which correspond to gravity modes. In the absence of a gradient in the mean molecular
weight, these reduce to the $g$-modes discussed in \citet{2010ApJ...720L..97B}.
The third root is given by
$\sigma\approx -a_3/a_2$, i.e.,
\be
\sigma&\approx& \tci \frac{N^2_{T\mu}}{N^2} \,,
\label{conduction_Bz}
\en
and corresponds to a mode driven by conduction.
Assuming that $N^2>0$, $g$-modes are overstable if the condition 
(\ref{RH-3.3}) is not satisfied, while conduction modes are unstable if
(\ref{RH-3.2}) is not fulfilled.

\begin{figure*}[]
\begin{center}
  \includegraphics[width=0.95\textwidth,trim=0 0 0 0]{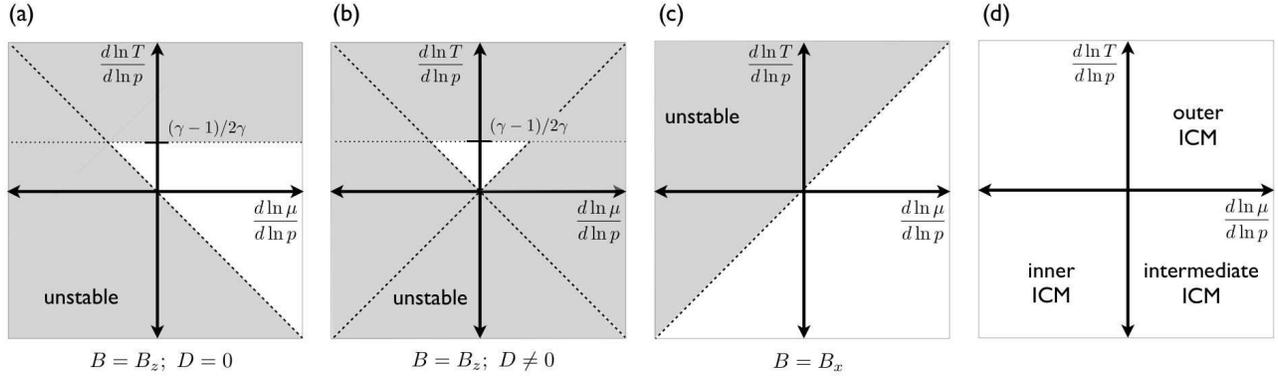}
  \caption{
  Panels $(a)$, $(b)$, and $(c)$ show a graphic representation of the stable 
  regions (white) satisfying all the Routh--Hurwitz stability criteria discussed in 
  Section~\ref{sec:modes}. Panel $(d)$ is a schematic representation of the correspondence 
  between the  regions of a representative galaxy cluster with radial temperature and mean molecular weight 
  profiles as shown in Figure~\ref{fig:f3} and the plane spanned by $(d\ln T/d\ln P, d\ln \mu/d\ln P)$.
}
  \label{fig:f2}
\end{center}
\end{figure*}

In order to understand how these modes behave in the parameter space spanned by
temperature and composition gradients, it is convenient to use that
$d\ln P/dz=-1/H$ and work with the dimensionless variables $d\ln T/d\ln P$
and $d\ln \mu/d\ln P$. The classical requirement for stability against buoyancy, i.e.,
$N^2 > 0$ becomes
\be
\label{TP-cond-1}
\frac{d\ln T}{d\ln P} < \frac{d\ln \mu}{d\ln P} + \frac{\gamma - 1}{\gamma} \,,
\en 
while the conditions (\ref{RH-3.2}) and (\ref{RH-3.3}) become, respectively,
\begin{eqnarray}
\label{TP-cond-2}
\frac{d\ln T}{d\ln P} &>& -\frac{d\ln \mu}{d\ln P} \,,  \\
\label{TP-cond-3}
\frac{d\ln T}{d\ln P} &<& \frac{\gamma - 1}{2\gamma} \,.
\end{eqnarray}
The first panel (a) in Figure~\ref{fig:f1} shows the regions of parameter space
where $g$-modes in Equation~(\ref{fast_overstable_Bz}) 
are stable, over-stable, or unstable (gray area).
The second panel (b) shows that the modes in Equation~(\ref{conduction_Bz}), 
which are driven by conduction, can be either stable 
or unstable. Note that in the region of parameter space where
both gravity and conduction modes are overstable/unstable they both grow with comparable rates. 
Panel (a) in Figure~\ref{fig:f2} shows the stable region (white) of parameter space satisfying
the conditions (\ref{TP-cond-2}) and (\ref{TP-cond-3}) simultaneously.

\subsection{Heat and Ion Diffusion Along Vertical Magnetic Field Lines}
\label{sec:Bz_Dneq0}

Let us now consider the situation in which ions
diffuse mainly along magnetic field lines.  If $D\ne0$, the dispersion
relation corresponding to Equations~(\ref{deltav}) to (\ref{deltamu})
yields as non-trivial roots $\sigma = \pm i\omega_{\rm A}$ and the
solutions to 
\be
\sigma^4+b_1\sigma^3+b_2\sigma^2+b_3\sigma+b_4=0 \,,
\label{pol_b} 
\en
with coefficients 
\begin{eqnarray}
b_1 &=& \tci + \tdi + \f{k_\bot^2}{k^2}\tvi \approx \tci \,, \nonumber \\
\label{b1}
b_2 &=&\f{k_\bot^2}{k^2}N^2+\omega_{\rm A}^2+ \tci\left(\tdi + \f{k_\bot^2}{k^2}\tvi\right) 
\approx \f{k_\bot^2}{k^2}N^2 \,, \\
b_3 &=& - \tci \left[\f{k_\bot^2}{k^2} N_{T\mu}^2 - \omega_{\rm A}^2\right]  + {\tdi}\left[\f{k_\bot^2}{k^2}(N^2 + \tci\tvi)+\omega_{\rm A}^2\right]  \nonumber  \\
&\approx& - \tci \f{k_\bot^2}{k^2} N_{T\mu} ^2 \,,\,\,\,\,\,\,\,\,\,\,\\
b_4 &=& -{\tdi}\tci\left[\f{k_\bot^2}{k^2}N_{T/\mu}^2-\omega_{\rm A}^2\right] 
\approx -{\tdi} \tci \f{k_\bot^2}{k^2}N_{T/\mu}^2  \,. 
\label{b4}
\end{eqnarray}
The approximations on each of the last terms on the right hand side 
hold under the same considerations that lead to the simplified 
coefficients $a_i$ in Section~\ref{sec:Bz_Deq0}. 
The approximate expressions for the coefficients 
$b_i$ and $a_i$ are identical for $i=1,2,3$.

The Routh--Hurwitz stability criteria for a quartic polynomial with
real coefficients requires: $b_1>0, b_4>0, b_1b_2-b_3>0$, and
$b_1b_2b_3-b_1^2b_4-b_3^2>0$.  While $b_1>0$ is trivially satisfied,
$b_4>0$ implies: 
\be 
\label{RH-4.2}
N^2_{T/\mu} < 0 \,.
\en
The other two conditions lead again to inequalities (\ref{RH-3.2}) and (\ref{RH-3.3}).
Therefore, the effects of diffusion require that only the additional condition 
(\ref{RH-4.2}) be met for the system to be stable.

To leading order, three of the roots of the dispersion relation
(\ref{pol_b}) are given by Equations~(\ref{fast_overstable_Bz}) and
(\ref{conduction_Bz}).  The fourth solution consists of a mode driven by ion diffusion
\be
\label{diffusion}
\sigma\approx-\tau^{-1}_{\rm d}\frac{N_{T/\mu}^2}{N_{T\mu}^2} \,.
\en
This is an unstable mode if either of the criteria 
(\ref{RH-3.2}) or (\ref{RH-4.2}) is unfulfilled.

Thus, if ions can diffuse along magnetic field lines, in addition to
requiring that the gradients in temperature, pressure, and mean molecular 
weight satisfy the inequalities (\ref{TP-cond-2}) and (\ref{TP-cond-3}), we must have
\begin{eqnarray}
\label{TP-cond-4}
\frac{d\ln T}{d\ln P} &>& \frac{d\ln \mu}{d\ln P} \,.
\end{eqnarray}

The third panel (c) in Figure~\ref{fig:f1} shows that the modes in Equation~(\ref{diffusion}), 
which are driven by diffusion, can be either stable
or unstable (gray). Their growth rates are estimated to be an order of magnitude
smaller than either $g$-modes or conduction modes. The importance of these diffusion modes 
resides in that they can become unstable in regions of parameter space which 
are stable against $g$-modes and conduction modes. 
Panel (b) in Figure~\ref{fig:f2} shows the stable region (white) of parameter space satisfying
all the conditions (\ref{TP-cond-2}), (\ref{TP-cond-3}), and (\ref{TP-cond-4}) simultaneously.

\subsection{Heat and Ion Diffusion Along Horizontal Magnetic Field Lines}
\label{sec:Bx_all}

If the background magnetic field is perpendicular to the thermal and composition
gradients, the dispersion relation that governs the stability of the atmosphere is
\footnote{In the absence of diffusion, the only result that is modified in this 
section is that the root $\sigma=-\tdi$ for $D\ne 0$ becomes $\sigma=0$ for $D=0$.} 
\be
\sigma^5+c_1\sigma^4+c_2\sigma^3+c_3\sigma^2+c_4\sigma+c_5=0 \,,
\label{pol_c}
\en
where the coefficients
\be
&&c_1\approx\tci+\tvi\f{k^2_\bot}{k^2}\,,\\
&&c_2\approx N^2 \f{k^2_x+k_y^2}{k^2}+\tci\tvi\f{k^2_\bot}{k^2}\,,\\
&&c_3\approx\tci N^2_{T/\mu}\f{k^2_x+k_y^2}{k^2}+\tvi N^2\f{k_y^2}{k^2}\,,\\
&&c_4\approx\tci\tvi N^2_{T/\mu}\f{k_y^2}{k^2}\,,\\
&&c_5\approx\omega_{\rm A}^2\tci N^2_{T/\mu}\f{k^2_x+k_y^2}{k^2} \,,
\en
are subject to the same considerations employed in deriving the approximate 
expressions for the coefficients $a_i$ and $b_i$. 

The Routh-Hurwitz stability criteria for a fifth degree polynomial requires
$c_1>0$, $c_1c_2-c_3>0$, 
$c_1c_2c_3-c_1^2c_4-c_3^2+c_1c_5>0$, 
$c_4(c_1c_2c_3-c_1^2c_4-c_3^2)>c_5(c_1c_2^2-2c_1c_4-c_2c_3+c_5)$, and  $c_5>0$.
In the limit under consideration, i.e., $\omega_{\rm dyn}\gg \tci > \tvi$, 
these inequalities become, respectively,
\be
&&\tci>0\,,\\
&&N^2-N^2_{T/\mu}>0\,,
\label{cond_c2}
\\
&&N^2_{T/\mu}(N^2-N^2_{T/\mu})>0\,,\\
&&(N^2_{T/\mu})^2(N^2-N^2_{T/\mu})>0\,,\\
&&N^2_{T/\mu}>0 \,.
\label{c5>0}
\en 
The first of these conditions is trivially satisfied. This is also the case for
inequality (\ref{cond_c2}), since it can be written as $[(\gamma-1)/\gamma][dP/dz]^2/P\rho>0$. 
Therefore, the only independent condition required for stability is $N^2_{T/\mu}>0$.

Two of the approximate solutions to the dispersion relation (\ref{pol_c}) are given by
$\sigma\approx\pm i c_2^{1/2}+(c_3-c_1c_2)/2c_2$, i.e.,
\be
\label{fast_overstable_Bx}
\sigma\approx\pm i\f{\sqrt{k_x^2+k_y^2}}{k}\sqrt{N^2}-\f{1}{2\tau_c}\left(1-\f{N^2_{T/\mu}}{N^2}\right) \,,
\en
which correspond to stable gravity modes. The other two solutions correspond to a conduction
and a viscous (decaying) mode, which are, respectively 
\be
\label{conduction_Bx}
\sigma\approx-\tci \frac{N^2_{T/\mu}}{N^2} \,,  \qquad
\sigma\approx-\tvi \frac{k_y^2}{k_x^2+k_y^2} \,.
\en

We conclude that when the magnetic field is perpendicular to 
the temperature and composition gradients, the stability of $g$-modes 
requires only that $N^2>0$, whereas the stability of conduction modes requires also
\be
\label{TP-cond-5}
\frac{d\ln T}{d\ln P} < \frac{d\ln \mu}{d\ln P}  \,.
\en 

The fourth panel (d) in Figure~\ref{fig:f1} shows the regions of parameter space
where $g$-modes in Equation~(\ref{fast_overstable_Bx}) 
are stable or unstable (gray).
The fifth panel (e) shows that the modes in Equation~(\ref{conduction_Bx}), 
which are driven by conduction, can be either stable 
or unstable. These regions are significantly different from the 
corresponding regions in panel (b), for which the direction of the background
magnetic field is parallel to the direction of the temperature and composition gradients.
Note that, unlike the case where $B=B_z$, $g$-modes and conduction modes
cannot be simultaneously unstable when $B=B_x$. 
Panel (c) in Figure~\ref{fig:f2} shows the stable region (white) of parameter space satisfying
the condition (\ref{TP-cond-5}).

\section{Astrophysical Implications}
\label{sec:discussion}

\begin{figure}[]
\begin{center}
  \includegraphics[width=0.475\textwidth,trim=0 10 0 -20]{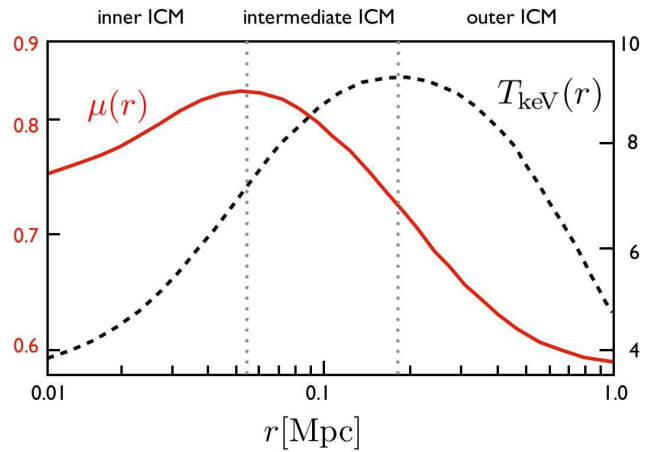}
  \caption{Schematic representation of the mean molecular weight (red solid line) and temperature (black dashed line)
  profiles of a representative galaxy cluster as suggested by observations \citep{2006ApJ...640..691V} 
  and theoretical models \citep{2011A&A...533A...6B}. The regions denoted by ``inner", ``intermediate", and ``outer" 
  ICM (delimited by dotted gray lines) correspond to three different quarters in the 
  $(d\ln T/d\ln P, d\ln \mu/d\ln P)$ plane (Fig.~\ref{fig:f2}.).
   The mean molecular weight for a homogeneous cluster with primordial abundance is $\mu \approx 0.59$.}
  \label{fig:f3}
\end{center}
\end{figure}

We have provided a detailed description of the various
instabilities that can be present in the parameter space spanned by
$(d\ln T/d\ln P, d\ln \mu/d\ln P)$ without imposing restrictions on
the relative signs of the gradients involved. 
We can now frame our results, summarized in Figure~\ref{fig:f2}, 
in the context provided by observations and theoretical models
addressing the temperature and composition structure of galaxy
clusters. 
Figure~\ref{fig:f3} shows a schematic representation of the
temperature and mean molecular weight profiles of a representative
galaxy cluster. The temperature profile sketched there resembles the
results obtained by observations \citep{2006ApJ...640..691V}, whereas
the mean molecular weight profile is akin to helium sedimentation
models \citep{2011A&A...533A...6B}.  If the peak in the temperature
takes place at a larger radius than the peak in the mean
molecular weight, then there are three distinct regions defined by the
signs of the temperature and composition gradients. These regions
correspond to different quadrants in the $(d\ln T/d\ln P, d\ln
\mu/d\ln P)$ plane as shown in panel (d) in Figure~\ref{fig:f2}.

The joint analysis of Figures~\ref{fig:f1} and \ref{fig:f2} allows us
to understand the implications that mean molecular weight gradients
can have for the various regions of a representative galaxy cluster 
as depicted in
Figure~\ref{fig:f3}.  If the magnetic field in the inner ICM is
perpendicular to the background temperature gradient $dT/dz>0$, as
suggested by the end states of initial configurations with $B=B_z$
that are HBI-unstable but evolve to $B\approx B_x$, then this region
is stable if $d\mu/dz = 0$ (although there could be overstable
$g$-modes driven by radiative cooling \citealt{2010ApJ...720L..97B}).
However, if $d\mu/dz >0$ this region is unstable to conduction modes.
If the magnetic field in the outer ICM is
parallel to the temperature gradient $dT/dz<0$, as suggested by the
end states of initial configurations with $B=B_x$ that are
MTI-unstable but evolve to $B\approx B_z$, then this region is
overstable to $g$-modes and unstable to conduction modes even if
$d\mu/dz = 0$.  A negative gradient in the mean molecular weight
is unable to stabilize these modes and it can further
drive unstable diffusion modes.  Therefore, even if the outer/inner ICM is stable
to both the MTI and the HBI, it could be rendered unstable to modes driven
by either gravity, conduction, or diffusion. Unless the magnetic field
is orthogonal to the background gradients, the region that
we have denoted as intermediate is likely to be unstable to
conduction and diffusion modes, but not to gravity modes. These $g$-modes, however, 
could be relevant if the temperature profile peaks at a smaller
radius than the mean molecular weight profile.

We will address the limit in which conduction is fast compared to the dynamical timescale
in a forthcoming paper. This limit contains the generalization of the instabilities 
that become the MTI and the HBI in the limit of a homogeneous plasma.

\newpage

\acknowledgements

We thank Matthew Kunz, Aldo Serenelli, and Shantanu Mukherjee for useful discussions.
MEP is grateful to the Knud H{\o}jgaard Foundation for its generous support.
SC acknowledges support from the Danish Research Council through FNU Grant 
No. 505100-50 - 30,168.

%\bibliography{ms}

\end{document}